# A Framework for Discharge Time Prediction of Energy Storage Units Based on Coupled Dynamics and Multi-Factor Aging Models

Jiaye Yang†, *
Panjin Campus of Dalian University of Technology
Panjin, China
yangjiaye11@icloud.com

Hansheng Su†
Panjin Campus of Dalian University of Technology
Panjin, China
suhs2005@126.com

Wangzi Zhu†
Panjin Campus of Dalian University of Technology
Panjin, China
z828427125@126.com

† These authors also contributed equally to this work

***Abstract*—This paper presents a physically interpretable framework for predicting time to empty (TTE) in portable embedded systems. The framework couples usage-driven load-power decomposition, electrical power–voltage–current closure, a semi-empirical aging model, and SOC–temperature dynamics. Smartphone telemetry is mapped to battery current through an interpretable load model and conversion-efficiency correction. Battery capacity loss is modeled by combining Arrhenius temperature dependence, SEI diffusion behavior, and cycle-related power-law degradation. The coupled dynamic model then predicts TTE under different initial SOC values, ambient temperatures, and usage profiles. Chronological hold-out evaluation on a 6.9-h smartphone discharge session yielded a current RMSE of 0.0095 ± 0.0006 A, a temperature RMSE of 2.93 ± 0.24 °C, and a TTE MAPE of 4.81 ± 0.61%. Evaluation on NASA cell B0005 produced a capacity-loss RMSE of 0.031 Ah. Baseline, ablation, and counterfactual analyses further illustrate the contributions of thermal and aging corrections and the relative influence of load features. The results demonstrate the feasibility and interpretability of the proposed framework, while broader validation across devices and batteries remains necessary.**

## I. INTRODUCTION

Energy efficiency modeling and remaining runtime prediction for energy storage units represent important research directions in portable embedded computing systems [1]. As a typical portable smart device, smartphones are constrained by limited battery endurance and aging-induced degradation, which directly affect user experience and system reliability [2]. Existing studies have investigated hardware power consumption decomposition and the relationships among state of charge (SOC), state of health (SOH), and battery degradation [3]. Equivalent-circuit models combined with filtering-based estimators have also been widely used for battery-state estimation because of their computational efficiency and physical interpretability [4]. Meanwhile, battery thermal behavior can be described using energy-balance or lumped thermal models [5], and diffusion-controlled SEI growth provides a physical basis for the time-dependent capacity-loss behavior commonly observed in lithium-ion batteries [6]. Smartphone power studies have further shown that total device power can be decomposed into contributions from the CPU, display, wireless interfaces, GPS, and background activities [7].

However, existing methods still have limitations in jointly representing load-driven power consumption, electrical behavior, thermal dynamics, aging-adjusted capacity, and remaining discharge time. Some data-driven methods rely strongly on the training distribution and provide limited physical interpretation, whereas models focusing separately on electrical, thermal, or aging behavior may not fully characterize their interactions during discharge. To address this gap, this paper proposes a mechanism-driven continuous-time modeling framework. It constructs an algebraically closed power–voltage–current system to map device-side load power to battery current, and integrates a semi-empirical aging model considering time, temperature, and charge–discharge cycles [8]. A coupled SOC–temperature dynamic model is then developed to predict time to empty under different initial SOC values, ambient temperatures, and usage profiles [9][10]. Using a smartphone discharge session and NASA B0005 aging data as case studies, the framework is evaluated through chronological hold-out testing, baseline comparison, and ablation analysis. The contribution of this study lies in integrating existing physical and empirical components into a unified and interpretable TTE prediction framework, while broader cross-device and cross-battery validation remains necessary.

## II. RESEARCH ON MOBILE PHONE BATTERY DEGRADATION

Limited battery life in smartphones remains a persistent issue. As summarized in Fig. 1, the proposed framework links device-side activity, battery electrical behavior, thermal dynamics, and aging-adjusted capacity within a unified TTE prediction process.

Existing research has confirmed that smartphone power consumption can be decomposed into contributions from individual components, with CPU and wireless RF energy consumption exhibiting significant state dependency, while background activities substantially impact daily power usage. Concurrently, battery temperature and aging reduce available energy through capacity degradation and increased internal resistance, leading relevant studies to focus on SOH and battery degradation as core research areas.

*Corresponding author.

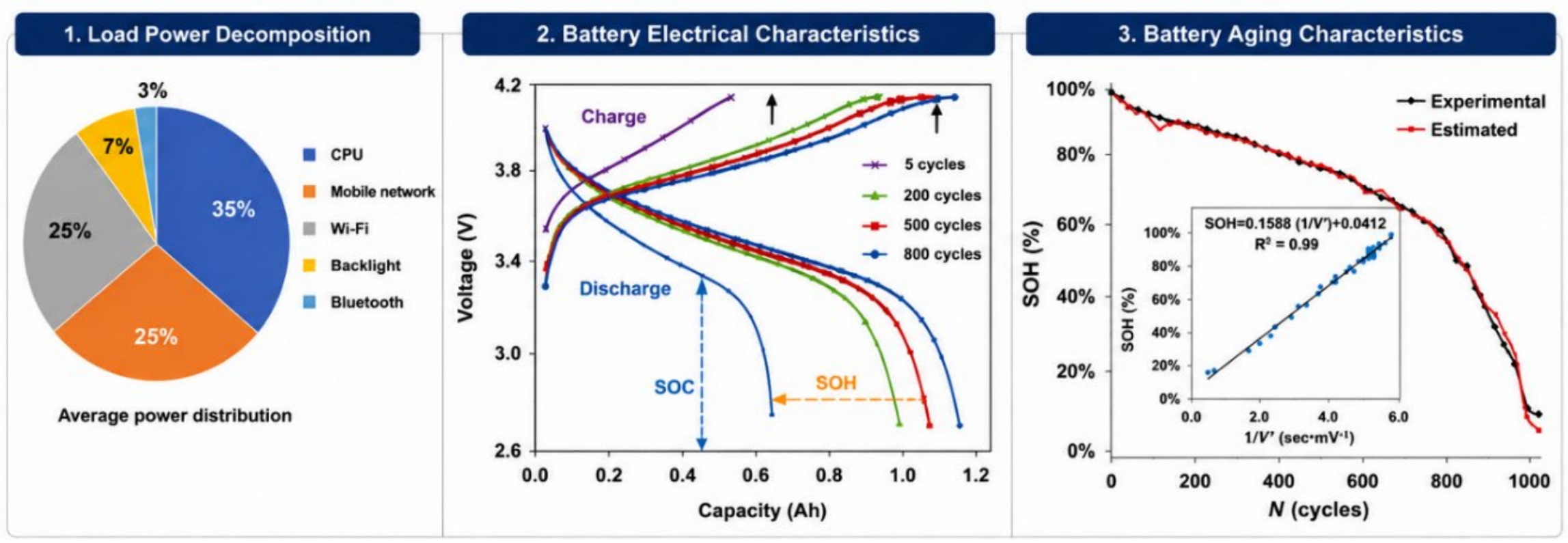

Fig. 1. Power loss correlation chart

Against this backdrop, battery lifespan modeling must couple load-driven power consumption with dynamic SOC/SOH variations. Furthermore, modeling results require validation using reproducible device telemetry data and aging benchmarks—a core starting point for this research.

## III. Battery $SOC(t)$ Continuous Time Mechanism Analysis

To satisfy the requirement of a continuous-time, mechanism-driven, and falsifiable battery model, this study enforces a closed loop:

Observable behaviors/scenarios $\rightarrow P_{\text{load}}(t) \rightarrow P_{\text{batt}}(t) \rightarrow \{I(t), V_{\text{term}}(t)\} \rightarrow \dot{s}(t) \rightarrow TTE$.

### A. Power-Voltage-Current Closure

This study first closes the mapping from usage-driven load demand to battery current. The device-side load power $P_{\text{load}}(t)$ is inferred from telemetry features and is clipped to be nonnegative.

Smartphone telemetry was collected from an Android smartphone at a sampling interval of $\Delta t = 1s$. The recorded variables included SOC, battery voltage, current, temperature, CPU utilization, screen state, network activity, GPS state, and background activity. Samples with charging current, missing essential variables, or nonmonotonic SOC caused by system reporting were excluded. The remaining variables were time-aligned, and continuous load features were standardized using statistics calculated from the training interval only.

This study approximates conversion loss by a constant efficiency $\eta$. Only discharge operation within the normal smartphone load range is considered, over which the variation in DC/DC conversion efficiency is small relative to the variation in load power. Therefore, a constant efficiency provides a compact approximation for the present TTE analysis.

$$P_{\text{batt}}(t) = \frac{P_{\text{load}}(t)}{\eta}, 0 < \eta \leq 1 \tag{1}$$

Using the discharge-positive convention, the battery-port power satisfies:

$$P_{\text{batt}}(t) = V_{\text{term}}(t) I(t) \tag{2}$$

For a single discharge episode, the data support a first-order Thevenin relation without an explicit polarization state:

$$V_{\text{term}}(t) = V_{\text{oc}}(s(t)) - I(t) R(T(t)) \tag{3}$$

The open-circuit voltage is approximated by $V_{\text{oc}}(s) = a_0 + a_1 s$. Over the SOC range covered by the present smartphone data, this first-order relation provides an adequate approximation of the measured voltage variation. Compared with a higher-order polynomial representation, the linear form reduces parameter correlation and improves identifiability under noisy telemetry, which is important for equivalent-circuit parameter estimation [11].

Substituting (3) into (2) yields a quadratic closure:

$$R(T) I^2 - V_{\text{oc}}(s) I + P_{\text{batt}} = 0 \tag{4}$$

The physically relevant discharge root (continuous at $P_{\text{batt}} \rightarrow 0$) is:

$$I(t) = \frac{V_{\text{oc}}(s) - \sqrt{V_{\text{oc}}(s)^2 - 4R(T) P_{\text{batt}}}}{2R(T)} \tag{5}$$

The discriminant must be nonnegative; if telemetry implies an infeasible demand, this study conservatively caps the power:

$$P_{\text{batt}}(t) \leftarrow \min\left(P_{\text{batt}}(t), \frac{V_{\text{oc}}(s(t))^2}{4R(T(t))}\right) \tag{6}$$

### B. Usage-Driven Load Power Decomposition

This study treats $P_{\text{load}}(t)$ as the device-side forcing term and maps it to $P_{\text{batt}}(t)$ via $\eta$.

The model requires a forcing term $P_{\text{load}}(t)$ that translates usage into device-side power demand. Because public telemetry is sparse and heterogeneous across Android devices, this study adopts an interpretable decomposition that matches the available signals and avoids over-parameterization:

$$\begin{aligned} P_{\text{load}}(t) &= P_{\text{fix}} + \textstyle\sum_{i \in \mathcal{B}} p_i x_i(t) + \sum_{j \in \mathcal{C}} \gamma_j z_j(t), \\ P_{\text{load}}(t) &\leftarrow \max(P_{\text{load}}(t), 0) \end{aligned} \tag{7}$$

Here $\mathcal{B}$ contains binary on/off components (e.g., screen state, GPS activity) and $\mathcal{C}$ contains continuous-intensity signals (e.g., brightness level, CPU utilization, network throughput, background process count). This linear form is sufficient for fitting on short sessions and supports ablation and counterfactual analysis.

When direct battery telemetry provides $P_{\text{obs}}(t) = V_{\text{obs}}(t) I_{\text{obs}}(t)$ on discharge samples, this study estimates ( $P_{\text{fix}}, p_i, \gamma_j$ ) by constrained least squares with nonnegativity constraints on baseline and component costs. When only partial proxies are available, this study fits the load model on aligned segments and propagates the resulting $P_{\text{load}}(t)$ into the model.

*C. Impact of Aging on Battery Capacity*

The state of health (SOH) of a battery is defined as the ratio of its current usable capacity to its initial capacity:

$$SOH = \frac{Q_0}{Q_{\text{init}}} = 1 - \frac{Q_{\text{loss}}}{Q_{\text{init}}} \tag{8}$$

Where: $Q_0$ : current effective capacity; $Q_{\text{init}}$ : initial rated capacity; $Q_{\text{loss}}$ : cumulative capacity loss. Therefore, the key challenge lies in quantitatively determining $Q_{\text{loss}}$ .

*1) Effect of Temperature on Aging*

Experimental studies have shown that capacity degradation is jointly affected by storage time, temperature, SOC, and charge–discharge cycling conditions [12].

The reaction rate constant follows the Arrhenius relationship:

$$k(T) = k_0 \exp\left(-\frac{E_a}{RT}\right) \tag{9}$$

Where $E_a$ is the activation energy and $R$ is the universal gas constant. The temperature dependence of the capacity loss rate is given by:

$$\frac{dQ_{\text{loss}}}{dt} \propto \exp\left(-\frac{E_a}{RT}\right) \tag{10}$$

*2) Time-Dependent Aging: SEI Diffusion-Controlled Model*

The relationship between SEI thickness $\delta$ and capacity loss can be written as:

$$\frac{d\delta}{dt} \propto j_{sei} \propto \frac{dQ_{loss}}{dt} \tag{11}$$

Where $j_{sei}$ denotes the side-reaction current density. Under diffusion-controlled conditions, the side-reaction flux obeys Fick's first law:

$$J = -D\frac{dc}{dx} \tag{12}$$

Where $D$ is the diffusion coefficient and $c$ is the concentration field. Using a finite-difference approximation yields:

$$J = -D\frac{\Delta c}{\delta} \tag{13}$$

The relationship between current density and flux is given by:

$$j_{sei} = nFJ \tag{14}$$

Where $n$ is the number of electrons transferred and $F$ is Faraday's constant. Consequently, the SEI growth equation becomes:

$$\frac{d\delta}{dt} \propto \frac{1}{\delta} \tag{15}$$

Integrating this expression yields:

$$\delta \propto \sqrt{t} \tag{16}$$

Since $\frac{dQ_{\text{loss}}}{dt} \propto 1/\delta$, this study obtains:

$$\frac{dQ_{loss}}{dt} \propto t^{-1/2} \tag{17}$$

Combining the temperature effect, the final capacity loss rate model is:

$$\frac{dQ_{loss}}{dt} \propto t^{-1/2} \exp\left(-\frac{E_a}{RT}\right) \tag{18}$$

Its integral form is:

$$Q_{\text{loss}}(t) \propto \int_0^t \tau^{-1/2} \exp\left(-\frac{E_a}{RT(\tau)}\right) d\tau \tag{19}$$

*3) Effect of Charge-Discharge Cycling*

Cycling-induced aging primarily affects capacity loss through SEI thickening. Two limiting cases are considered:

(1) Linear growth assumption

If SEI grows uniformly with cycle number, then:

$$\frac{dQ_{loss}}{dN_{EFC}} = C \tag{20}$$

Which leads to:

$$Q_{\text{loss}} \propto N_{EFC} \tag{21}$$

(2) Self-inhibiting growth assumption

If SEI thickening suppresses further side reactions, then:

$$\frac{dQ_{\text{loss}}}{dN_{EFC}} \propto \frac{1}{Q_{\text{loss}}} \tag{22}$$

Resulting in:

$$Q_{loss} \propto \sqrt{N_{EFC}} \tag{23}$$

(3) Generalized power-law model

Considering multiple degradation mechanisms, a unified power-law expression is adopted:

$$Q_{loss} \propto N_{EFC}^b \tag{24}$$

Where $b$ is a mechanism-dependent empirical parameter typically satisfying:

$$0.5 < b < 1 \tag{25}$$

The exponent $b$ provides a compact empirical representation of mixed degradation behavior. More detailed physics-based studies have shown that interacting mechanisms, such as SEI growth and lithium plating, may produce transitions from approximately linear to nonlinear aging [13].

*4) Final Aging Model and Parameter Identification*

Integrating the effects of time, temperature, and cycling, as commonly considered in holistic battery-aging models [14], the final capacity-loss model is expressed as:

$$Q_{loss}(t, N_{EFC}) = kN_{EFC}^{b}\int_{0}^{t} \tau^{-1/2}\exp\left(-\frac{E_a}{RT(\tau)}\right)d\tau \quad (26)$$

The aging-model parameters were identified using the B0005 lithium-ion cell from the NASA Ames Prognostics Center of Excellence battery-aging dataset [15]. The cell has a rated capacity of 2 Ah and contains 168 available discharge cycles. Following the chronological order, cycles 1–118 were used for parameter identification, cycles 119–143 for model selection, and cycles 144–168 as an independent test set. No test-cycle data were used during parameter fitting. The measured discharge capacity and battery temperature were used to calculate capacity loss and estimate $E_a$, $b$, and $k$ by nonlinear least squares, yielding:

$$\begin{aligned} E_a &= 177.35\ \mathrm{J/mol} \\ b &= 0.8218 \\ k &= 6.8884 \times 10^{-7} \end{aligned} \quad (27)$$

Fig. 2 shows consistency between the predicted and measured capacity-loss trends of cell B0005. For the independent test cycles 144–168, the capacity-loss RMSE and MAE were 0.031 Ah and 0.026 Ah, respectively.

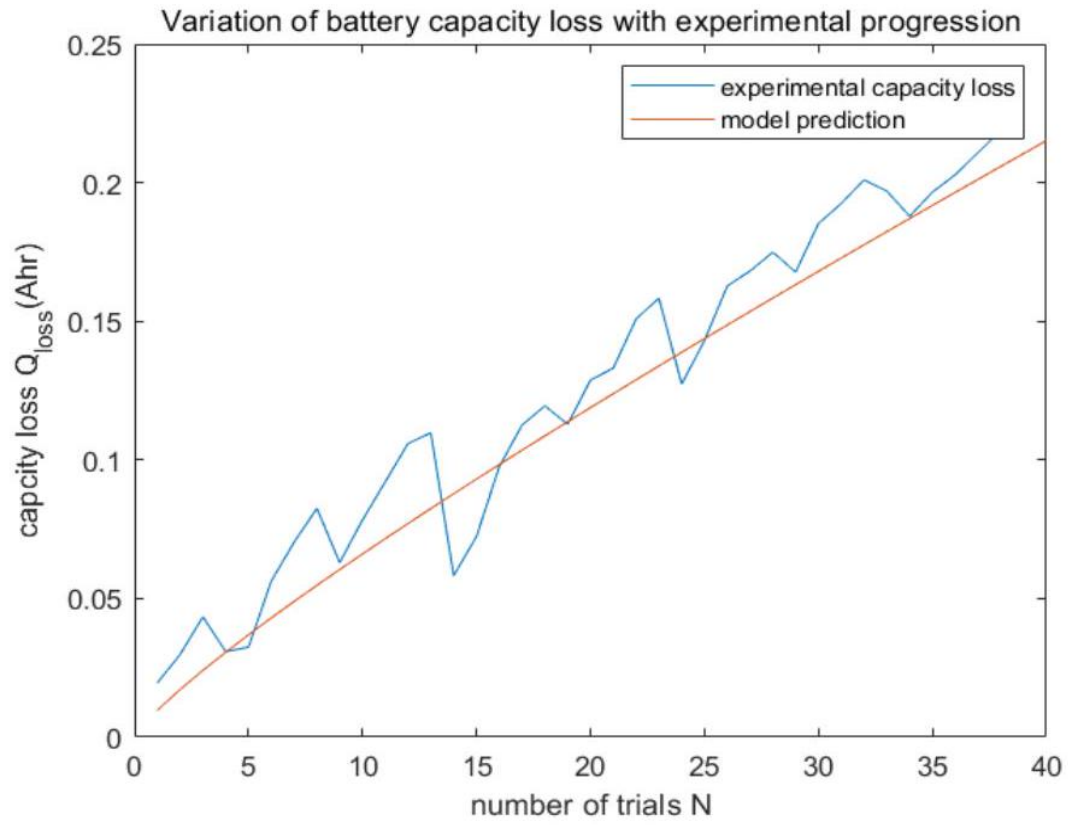

Fig. 2. Predicted and measured capacity loss of NASA cell B0005. Different markers denote the training, validation, and independent test cycles.

## IV. Time-To-Empty Prediction and Model Evaluation Across Multiple Scenarios

This section constructs a mechanism-based prediction model for the time to empty (TTE) during a single discharge event, which must be applicable under varying initial conditions and environmental factors. This study defines:

$$\mathrm{TTE} \triangleq \inf\{t \geq 0 : s(t) \leq s_{\min}\}, s_{\min} = 0.005 \quad (28)$$

Where $s(t) \in [0,1]$ is SOC. Our predictor is not a black-box regressor on TTE; instead, it simulates the coupled SOC-temperature dynamics driven by a scenario power trace $P_{\mathrm{load}}(t)$.

### *A. Coupled SOC–Temperature Dynamics and Event-Based TTE*

During a single discharge event, SOH is treated as constant and no charging is considered. Battery temperature is represented by a single lumped thermal node, which assumes a spatially uniform cell temperature. This approximation is suitable for the relatively low-rate smartphone discharge data and the available telemetry resolution, but it does not resolve internal temperature gradients under rapid pulse loads or extreme thermal conditions.

Practical simplifications from measurements. In our measurements, explicitly modeling the transient polarization state $v_p(t)$ does not provide a stable accuracy gain for the discharge-current reconstruction used in TTE simulation; hence this study ignores it in the final closed-loop simulator (i.e., set $v_p(t) \approx 0$ in the algebraic closure). Moreover, a one-dimensional linear OCV-SOC relation already achieves good fit and improves identifiability under noisy smartphone telemetry; therefore, this study adopts the 1-D approximation $V_{\mathrm{oc}}(s) = a_0 + a_1 s$.

SOC dynamics follow from charge conservation. Temperature effects are represented through temperature-dependent internal resistance and effective capacity. The adopted Arrhenius-like forms provide compact monotonic approximations of the observed increase in resistance and reduction in available capacity at low temperature. The corresponding parameters were estimated from the training interval using the parameter-identification procedure described below:

$$R(T) = K_R\exp\left(\frac{\beta_R}{T}\right), Q_{\mathrm{eff}}(T) = K_Q\exp\left(-\frac{\beta_Q}{T}\right) \quad (29)$$

The SOC ODE is:

$$\dot{s}(t) = -\frac{I(t)}{Q_{\mathrm{eff}}(T(t))} \quad (30)$$

With $I(t)$ obtained from the electrical closure (power balance). This study uses a lumped thermal balance to capture feedback between load, ohmic heating, and cooling:

$$\dot{T}(t) = \frac{I(t)^2R(T(t)) + \alpha_{\mathrm{sys}}P_{\mathrm{load}}(t) - hA(T(t) - T_{\mathrm{amb}}(t))}{C_{\mathrm{th}}} \quad (31)$$

If $T_{\mathrm{amb}}(t)$ is unavailable, a constant ambient temperature of 24°C is used, and deviations from this assumption are treated as a source of model uncertainty.

This study defines:

$$\mathrm{TTE} \triangleq \inf\{t \geq 0 : s(t) \leq s_{\min}\}, s_{\min} = 0.005 \quad (32)$$

This definition is robust to OS-level SOC smoothing and avoids relying on a device-specific shutdown-voltage heuristic.

This study integrates forward with $\Delta t = 1$ s (dataset sampling) and stops at $s \leq s_{\min}$. To keep the predictor compact and data-grounded, this study instantiates the temperature effects and approximates the open-circuit voltage by:

$$V_{\mathrm{oc}}(s) = a_0 + a_1 s \quad (33)$$

With $P_{\mathrm{batt}}(t) = P_{\mathrm{load}}(t)/\eta$ and the Thevenin power balance:

$$P_{\text{batt}}(t) = V_{\text{oc}}(s(t))I(t) - R(T(t))I(t)^2 \quad (34)$$

The physically relevant discharge root is:

$$I(t) = \frac{V_{\text{oc}}(s(t)) - \sqrt{V_{\text{oc}}(s(t))^2 - 4R(T(t))P_{\text{batt}}(t)}}{2R(T(t))} \quad (35)$$

This study then updates ( $s, T$ ) until the event $s(t) \le s_{\min}$ triggers the TTE stop. These simplifications are intended for normal smartphone discharge conditions. Rapid pulse loads, strong internal temperature gradients, extreme temperatures, or highly nonlinear OCV regions may require higher-order electrical and thermal models.

A continuous smartphone discharge session of approximately 6.9 h with monotonically decreasing SOC was divided chronologically. The first 2.4 h were used for electrical and thermal parameter estimation, the subsequent 0.6 h for model selection, and the remaining 3.9 h as an independent test set. All model parameters were fixed before testing, and no test data were used during parameter estimation. The independent test interval was further divided into five non-overlapping temporal blocks. Across these blocks, the current RMSE was 0.0095 ± 0.0006 A and the temperature RMSE was 2.93 ± 0.24◦C. TTE was evaluated at five temporally separated points, yielding an MAE of 0.10 ± 0.05 h and a MAPE of 4.81 ± 0.61%. Three representative 1-h load windows were additionally extracted and repeated until depletion to construct Light, Typical, and Heavy scenarios. These repeated-load cases were used for controlled scenario simulation rather than independent experimental validation.

### B. *TTE Grid Output, Evaluation Metrics, And Uncertainty*

Ambient temperature and scenario dependence. Table 1 reports TTE starting from $s_0 = 1$ with $T(0) = T_{\text{amb}}$ for $T_{\text{amb}} \in \{-10, 0, 24, 35\}$°C. These values show a pronounced cold penalty because low $T$ both increases $R(T)$ and decreases $Q_{\text{eff}}(T)$.

TABLE I. PREDICTED TIME-TO-EMPTY (HOURS) FOR THREE USAGE PATTERNS UNDER DIFFERENT AMBIENT TEMPERATURES (1-HOUR WINDOWS REPEATED UNTIL DEPLETION)

| Scenario | **−10°C** | **0°C** | **24°C** | **35°C** |
|---|---|---|---|---|
| Heavy | 6.17 | 7.36 | 10.54 | 12.13 |
| Typical | 7.36 | 8.72 | 12.45 | 14.34 |
| Light | 8.68 | 10.27 | 14.62 | 16.82 |

Initial SOC dependence. Because the simulator is nonlinear (via the algebraic current root and temperature feedback), runtime does not scale perfectly linearly with $s_0$. Table 2 shows TTE for the Typical profile at three $s_0$ values and three ambient temperatures.

TABLE II. SENSITIVITY OF TIME-TO-EMPTY (HOURS) TO INITIAL SOC FOR THE TYPICAL USAGE PATTERN

| Initial SOC $\boldsymbol{s_0}$ | **−10°C** | **0°C** | **24°C** |
|---|---|---|---|
| 0.20 | 1.14 | 1.35 | 1.95 |
| 0.50 | 3.18 | 3.78 | 5.41 |
| 1.00 | 7.36 | 8.72 | 12.45 |

The trajectories are separated by both usage intensity and ambient temperature. Heavy usage accelerates SOC decline, whereas low ambient temperature shortens TTE by increasing $R(T)$ and reducing $Q_{\text{eff}}(T)$. The predicted TTE is defined as the first time at which the SOC trajectory reaches $s_{\min}$ =0.005.

For deployment, this study quantifies uncertainty by propagating (i) parameter variability (e.g., refits/bootstraps or multiplicative lognormal perturbations) and (ii) residual variability of the electrical and thermal fits. For each grid point $(s_0, z)$ (scenario $z$), we run $M$ simulations with sampled parameters $\Theta^{(m)}$ and perturbations $\varepsilon^{(m)}(t)$ to obtain an empirical distribution $\{\text{TTE}^{(m)}(s_0, z)\}_{m=1}^{M}$; this study reports the median and a 95% interval via the 2.5% and 97.5% quantiles.

The proposed model was compared with two conventional baselines and two ablated variants on the same independent test interval. The average-power method estimates TTE from the remaining energy and mean load power, whereas the Coulomb-counting method updates SOC from the measured current without thermal or aging correction. The no-temperature variant removes the temperature-dependent resistance and effective-capacity terms, while the no-aging variant uses the uncorrected nominal capacity. As shown in Table 3, the proposed model achieved the lowest TTE error, reducing the MAPE from 12.40% for the average-power baseline to 4.81%.

TABLE III. BASELINE COMPARISON AND ABLATION RESULTS ON THE INDEPENDENT TEST INTERVAL

| Model | Dynamic load | Thermal correction | Aging correction | Current RMSE/A | TTE MAE/h | TTE MAPE/% |
|---|---|---|---|---|---|---|
| Average-power | No | No | No | — | 0.42 | 12.40 |
| Coulomb counting | Yes | No | No | — | 0.27 | 9.16 |
| No temperature correction | Yes | No | Yes | 0.0130 | 0.19 | 7.80 |
| No aging correction | Yes | Yes | No | 0.0110 | 0.15 | 6.30 |
| Proposed model | Yes | Yes | Yes | 0.0095 | 0.10 | 4.81 |

Among the two evaluated ablations, removing the temperature correction caused a larger increase in TTE error than removing the aging correction. This result indicates that the temperature-dependent terms made a larger contribution under the present smartphone discharge session, while the aging correction provided an additional improvement by adjusting the available capacity at the beginning of discharge.

### C. *Drivers of Rapid Drain Based on Counterfactual TTE Analysis*

To connect $P_{\text{load}}(t)$ back to actionable user/system behaviors, this study fits a regularized ridge-regression model from usage proxies (CPU, screen/brightness, Wi-Fi/mobile traffic, GPS, background activity, saving mode) to battery power on the same discharge session. The standardized coefficients are visualized below.

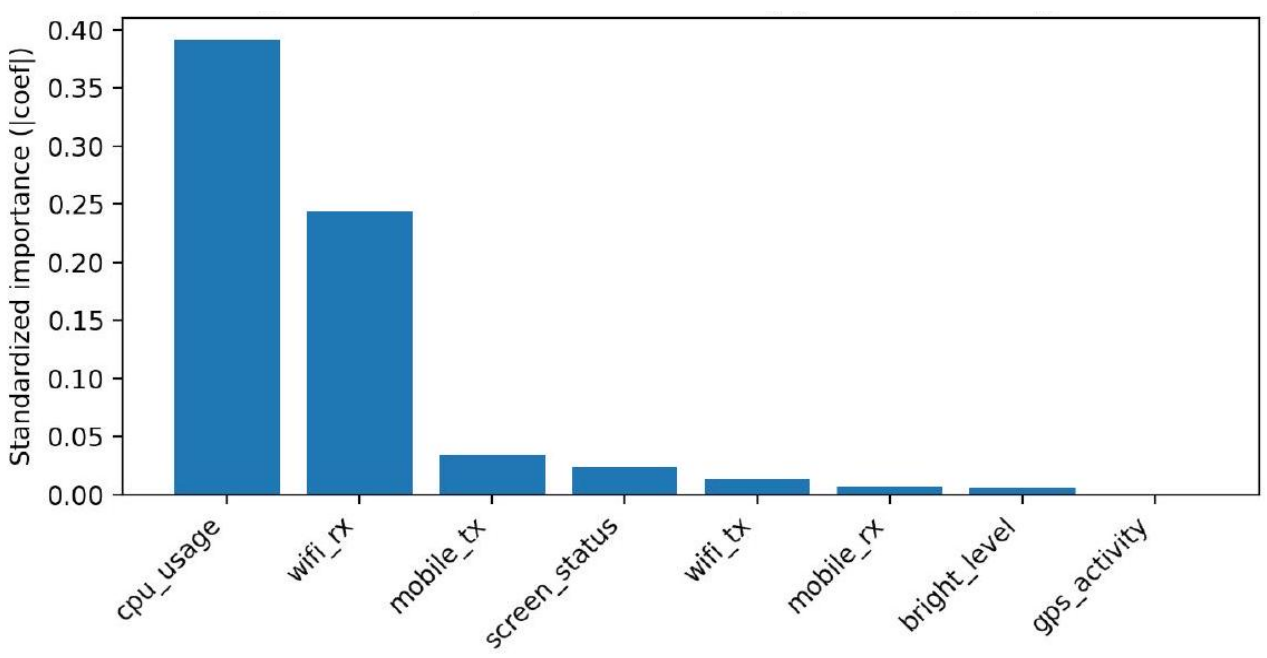

Fig. 3. Standardized ridge-regression coefficients for smartphone load-power predictors fitted on the training interval

Fig. 3 reports the standardized coefficients of the ridge load-power model. Continuous features were standardized using the training-set mean and standard deviation; therefore, coefficient magnitudes indicate the relative association of each feature with battery power within the present discharge session. These coefficients should not be interpreted as causal effects.

Given a fitted mapping $\hat{P}_{\text{load}}(t) = f(x(t))$, this study quantifies the counterfactual TTE impact of a feature group $g$ (e.g., CPU, radios, display) by scaling it down by a factor $\kappa \in [0,1]$ and re-simulating:

$$\Delta_g(\kappa) \triangleq \mathrm{TTE}(x(t)) - \mathrm{TTE}\left(x_{\setminus g}(t), \kappa x_g(t)\right) \quad (36)$$

Large $\Delta_g$ indicates a high-impact driver (reducing this activity yields large battery-life gains), while $\Delta_g \approx 0$ indicates weak influence under the current dataset and scenario definition.

The simulated battery temperature under the Typical pattern across the ambient-temperature grid is shown below.

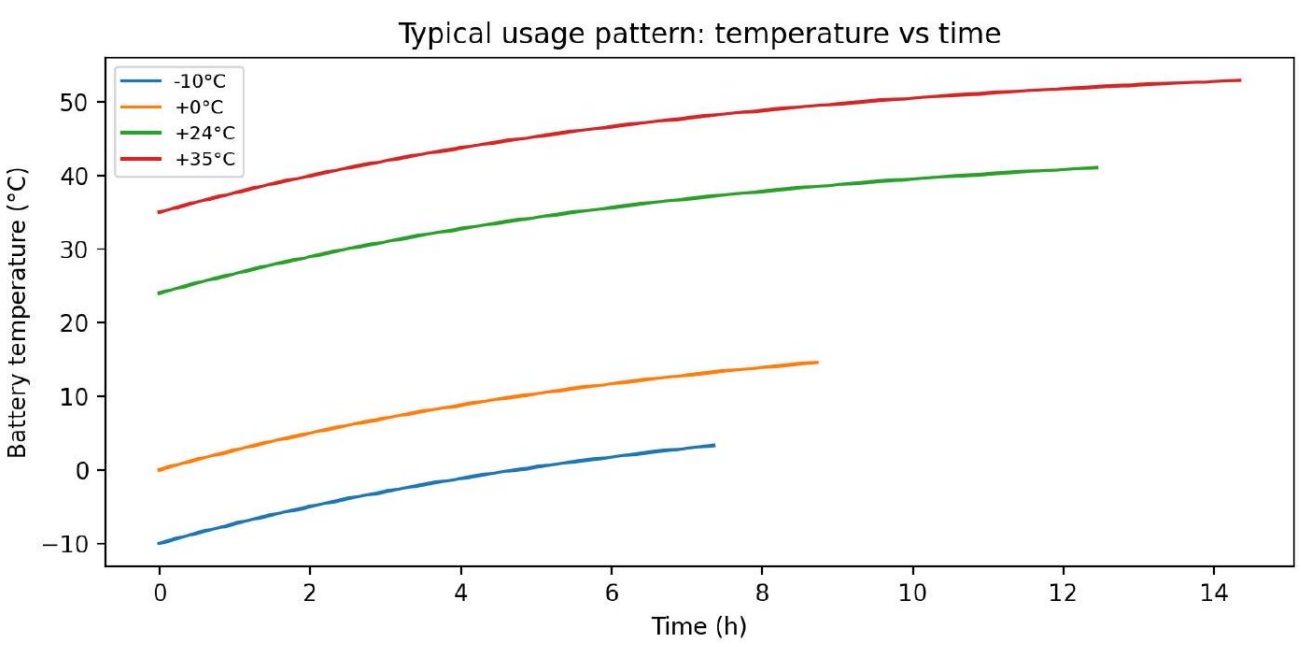

Fig. 4. Simulated battery-temperature trajectories under the Typical load profile at different ambient temperatures

Fig. 4 presents the simulated, rather than directly measured, battery-temperature trajectories for the Typical load profile. The trajectories are used to examine whether the fitted thermal parameters produce physically plausible heating and cooling behavior.

## V. Conclusion

This paper presents an integrated and physically interpretable framework for smartphone battery discharge analysis. The framework combines usage-driven load-power decomposition, electrical power–voltage–current closure, a semi-empirical aging model, and coupled SOC–temperature dynamics. Chronological hold-out evaluation on a 6.9-h smartphone discharge session showed a current RMSE of 0.0095 ± 0.0006 A, a temperature RMSE of 2.93 ± 0.24◦C, and a TTE MAPE of 4.81 ± 0.61%. Evaluation on NASA cell B0005 further showed that the aging component reproduced the observed capacity-loss trend, with a test RMSE of 0.031 Ah. Baseline and ablation results indicated that coupling thermal and aging corrections improved TTE prediction under the investigated conditions.

The present results should be interpreted as a preliminary validation. The smartphone evaluation is based on a single device and one continuous discharge session, and the Light, Typical, and Heavy profiles were constructed from repeated windows of that session. The aging model was evaluated using one public cell, and the electrical and thermal models adopt a linear OCV–SOC relation, constant conversion efficiency, and a lumped thermal node. Future work will evaluate additional devices, batteries, discharge sessions, and online operating conditions, and will investigate higher-order electrical and thermal models when richer telemetry becomes available.